\documentclass[fleqn,usenatbib]{mnras}

\usepackage{newtxtext,newtxmath}

\usepackage[T1]{fontenc}

\DeclareRobustCommand{\VAN}[3]{#2}
\let\VANthebibliography\thebibliography
\def\thebibliography{\DeclareRobustCommand{\VAN}[3]{##3}\VANthebibliography}

\usepackage{graphicx}	
\usepackage{amsmath}	

\title[X-ray and UV Variability of Quasars]{The Relation between X-ray and Ultraviolet Variability of Quasars}

\author[H. Sou et al.]{
Hao Sou,$^{1, 2}$\thanks{E-mail: Houston@mail.ustc.edu.cn}
Jun-Xian Wang,$^{1, 2}$\thanks{E-mail: jxw@ustc.edu.cn}
Zhang-Liang Xie,$^{3}$ 
Wen-Yong Kang, $^{1, 2}$
Zhen-Yi Cai, $^{1, 2}$
\\
$^{1}$CAS Key Laboratory for Research in Galaxies and Cosmology, Department of Astronomy, University of Science and Technology of
China, Hefei, Anhui 230026, China\\
$^{2}$School of Astronomy and Space Science, University of Science and Technology of China, Hefei 230026, China\\
$^{3}$Max Planck Institut für Astronomie, Königstuhl 17, D-69117, Heidelberg, Germany
}

\date{Accepted XXX. Received YYY; in original form ZZZ}

\pubyear{2015}

\begin{document}
\label{firstpage}
\pagerange{\pageref{firstpage}--\pageref{lastpage}}
\maketitle

\begin{abstract}
	The relation between X-ray and UV/optical variability in AGNs has been explored in many individual sources, 
	however a large sample study is yet absent. 
	Through matching the {\em XMM-Newton} serendipitous X-ray and UV source catalogs with SDSS quasars, we build a sample 
	of 802 epoch-pairs of 525 quasars showing clear variability in $ {\rm log}F_{\rm X} - {\rm log}F_{\rm UV} $ space. 
	After correcting for the effect of photometric noise, we find $35.6\pm2.1\%$ of the epoch-pairs show 
	asynchronous variability between X-ray and UV (brightening in one band but dimming in the other). 
	This indicates only in $28.8\pm4.2\%$ of the epoch-pairs the X-ray and UV variability are intrinsically coordinated. 
	The variability synchronicity exhibits no dependence on physical parameters of quasars or the time lag of the epoch-pairs, 
	except for stronger variability tends to have stronger synchronicity. Switches between synchronous 
	and asynchronous variability are also seen in individual sources. 
	The poor coordination clearly contradicts both the X-ray reprocessing model and the accretion rate variation model for AGN variability.
	The ratios of the observed X-ray variability amplitude to that in UV span a broad range and peak at $\sim$ 2.
	The dominant fraction of the ratios appear too small to be attributed to X-ray reprocessing, 
	and too large for accretion rate variation. 
	The inhomogeneous disk model which incorporates both X-ray and UV/optical variability in AGNs is favored by the observed stochastic relation between X-ray and UV variations.  
\end{abstract}




\section{Introduction}
\label{sec:intro}

The multi-band variability is one of the intriguing properties showing the active characteristics
of active galactic nuclei (AGN).
To explore the variability in the innermost region of AGNs, one generally focuses on the intrinsic emission from the central engine, i.e., 
the UV/optical radiation from the accretion disk \citep{1973A&A....24..337S,2010ApJ...712.1129M}, 
and the X-ray emission originated from the hot corona \citep{1991ApJ...380L..51H} through inverse Compton scattering
the thermal seed photons of the disk. In general the X-ray variability is more rapid than that in UV/optical, showing the corona is more compact than the disk region responsible for UV/optical emission. 

Tight inter-band correlations of UV/optical continuum variability, together with inter-band lags, have been detected by monitoring campaigns of individual AGNs \citep[e.g.,][]{1991ApJ...371..541K,1996ApJ...470..364E,2015ApJ...806..129E,2017ApJ...840...41E,1992ApJ...393..113C,2002MNRAS.332..231U,2016ApJ...821...56F}. The reprocessing model \citep{1991ApJ...371..541K}, in which UV/optical variations are attributed to reprocessed emission from the accretion disk, which is illuminated by a central variable source (presumably the X-ray corona), has been widely adopted to explain observations \citep[e.g.,][]{2012MNRAS.422..902C,2014ApJ...788...48S, 
2016ApJ...828...78N, 2020MNRAS.494.4057P}.
However note a recent sample study of quasars has revealed that, while the UV and optical flux variations could be fully correlated in $\sim$ 60\% of sources, they appear uncorrelated at all in the rest $\sim$ 40\% of quasars \citep{2020MNRAS.495.1403X}.

The correlation between X-ray and UV/optical flux variability in AGNs, though also detected in some individual AGNs \citep[e.g.,][]{1996ApJ...470..364E,2017ApJ...840...41E,1992ApJ...393..113C,2002MNRAS.332..231U, 2008MNRAS.389.1479A}, appears much weaker than that between UV and optical \citep[e.g.,][]{2019ApJ...870..123E, 2021PASA...38...42K}. Particularly in many case studies no correlation between X-ray and UV/optical flux variations could be detected at all \citep[e.g.,][]{1990MNRAS.243..713D,2000ApJ...544..734N, 2002AJ....124.1988M, 2017MNRAS.464.3194B, 2018MNRAS.475.2306B, 2018MNRAS.478.2557G, 2019ApJ...870...54M}, clearly challenging the simple X-ray reprocessing scheme (see \S\ref{sec:discussion} for other challenges to the reprocessing model).
Contrarily, the inhomogeneous disk model for AGN variability \citep{2011ApJ...727L..24D, 2016ApJ...826....7C,2018ApJ...855..117C,2020ApJ...892...63C}, which could naturally produce uncorrelated variation between different bands, would be favored (see \S\ref{sec:discussion} for further discussion).

Comparing the variability amplitude in X-ray to that in UV/optical is also helpful to test the X-ray reprocessing scheme
\citep[e.g.,][]{2003ApJ...584L..53U,2006ASPC..360..101U,2007ASPC..373..596G, 2009MNRAS.397.2004A}. In the X-ray reprocessing scheme, the variability amplitude of the UV/optical emission is expected to be much weaker than that in X-ray, unless the variable X-ray emission dominates the whole energy budget of the central engine. An example is that \cite{2003ApJ...584L..53U} found the optical variability amplitude in NGC 5548 is even larger than that in X-ray, severely against the simple X-ray reprocessing scheme. 

In literature, another widely adopted model for AGN variability is changes in accretion rate \citep[e.g.,][]{Pereyra_2006, 2018FrASS...5...19B}, although the timescale of changes in global accretion rate by viscosity is expected to be $10^{3-5}~$years \citep{1985ApJ...288..205A}.
As stronger emission in both X-ray and UV/optical are expected when the accretion rate increases, observing the correlation between X-ray and UV/optical variability could also provide essential test to this model. Furthermore, it is known that AGNs with higher Eddington ratios tend to have smaller X-ray to bolometric luminosity ratios \citep[e.g.,][]{2010ApJS..187...64G,2012MNRAS.425..623L}.
As pointed out by \cite{Wu2020}, if individual AGNs show stronger long-term variability in X-ray compared with UV/optical (thus larger X-ray to bolometric luminosity ratio at brighter phases), such variations can not be attributed to changes in accretion rate. 

However, because of requiring simultaneous observations from different instruments and at multiple epochs (X-ray: space-based; and UV/optical: space or ground-based), studies on the relation between X-ray and UV/optical flux variation (i.e., the correlation, and comparing the variability amplitudes) of a statistically large sample of AGNs have been absent.
In this work we represent a statistical analysis of the relation between X-ray and UV variability in a large sample of AGNs observed by {\em  XMM-Newton} at multiple epochs, thanks to the XMM-Newton Optical Monitors which provides simultaneous UV/optical coverage to X-ray. 
We focus on quasars in this work, as for which the host contamination to UV photometry could be negligible.

In \S\ref{sec:selection}, we elaborate our
quasar sample selection and data collection. 
\S\ref{sec:result} provides our primary
statistical results. 
We discuss and compare the result with predictions from existing models in \S\ref{sec:discussion}. 
Throughout the paper, we adopt the cosmology of
$H_{\rm 0} = 70{\rm ~km~s^{-1}~Mpc^{-1}}$, $(\Omega_{\rm M}, \Omega_{\rm \Lambda}) = (0.3, 0.7)$.
The ``log'' symbol denotes the base-10 logarithm.

\section{Sample Selection and methodology} 
\label{sec:selection}

The {\em XMM-Newton} Serendipitous Source Catalog (XMM-SSC) has been updated to the
fourth-generation, data release 10 (4XMM DR10, \citealt{2020A&A...641A.136W}),
containing 849,991 detections of 575,158 unique X-ray sources with 11,647 observation epochs.
The {\em XMM-Newton} Optical Monitor (XMM-OM, \citealt{2001A&A...365L..36M}) provides simultaneous UV/optical observations, 
and the latest {\em XMM-Newton} Serendipitous Ultraviolet Source Survey catalog (XMM-SUSS5, \citealt{2021yCat.2370....0P}) consists of 8,863,922 detections of
 5,965,434 sources with 10,628 observation epochs.

We cross-match the XMM-SSC and XMM-SUSS catalogs within 2\arcsec\ (selected
according to the X-ray position accuracy for sources at large off-axis angles).
For X-ray detections we adopt the EPIC X-ray fluxes from XMM-SSC (weighted mean fluxes from PN, MOS1 and MOS2) in the 0.5 -- 4.5~keV band (the XID band, short for X-ray follow-up and identification program, \citealt{2001A&A...365L..51W}), with the 4.5~keV threshold imposed to avoid high energy background and low instrument throughput. 
We drop a few null fluxes due to poor point spread function (PSF) coverage. Hereafter throughout this paper, $F_X$ (or log$F_X$, $\Delta$log$F_X$) refers to XMM-SSC EPIC 0.5 -- 4.5 keV flux, unless otherwise stated (see \S\ref{sec:discussion} for discussion).

For UV data, we focus on source detections in UVW1 (with an effective wavelength of 291~nm and a width of 93~nm), 
which has much better throughput compared with other UV filters (UVW2 at 212~nm, and UVM2 at 231~nm). Moreover, UVW1 band provides the most non-null fluxes compared to other bands (including U and B).
We then select the data with the same ``OBS\underline{~~}ID'' parameter to acquire
the simultaneity of X-ray and UV observations.

Secondly, we match (within 2\arcsec) XMM sources based on the UV positions 
with the Sloan Digital Sky Survey (SDSS) Quasar Catalog Data Release 16 
of the extended Baryon Oscillation Spectroscopic Survey \citep[eBOSS,][]{2020ApJS..250....8L}.
The selection yields a sample of 762 quasars with multi-epoch ($\ge$ 2) XMM 0.5 -- 4.5 keV and UVW1 fluxes. 
By grouping each two consecutive observation epochs of individual sources with multiple epochs as an epoch-pair, 
we obtained a total number of 1282 epoch-pairs.
Note for quasars with $N > 2$ epochs, we only consider $N-1$ pairs of two consecutive epochs, to avoid repeated use of epochs which could yield epoch-pair(s) not independent of others. For example, a quasar with 3 epochs ordered in time, we only consider the pair of epoch 1 and 2, and of epoch 2 and 3. We do not consider the pair of epoch 1 and 3, as this pair is not independent of the two pairs we adopted.

\begin{figure}
    \centering
    \includegraphics[width=\columnwidth]{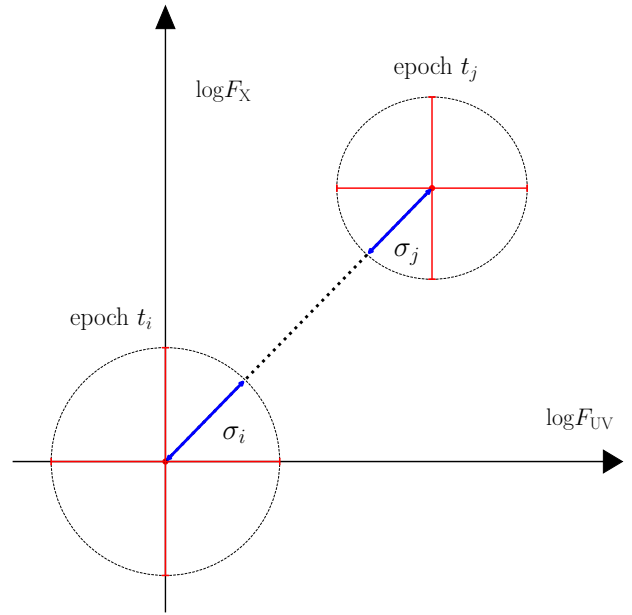}
    \caption{Schematic of selecting epoch-pairs with significant variation in $ {\rm log}F_{\rm X} - {\rm log}F_{\rm UV} $ space. 
	The horizontal and vertical semi-axes of the ellipses denote the uncertainties in ${\rm log}F_{\rm UV}$ and ${\rm log}F_{\rm X}$ of the corresponding epoch, respectively. 
	The variation of an epoch-pair is depicted by the distance between the pair in 2D space (dotted line), with an uncertainty $\sigma = \sqrt{\sigma_{i}^{2} + \sigma_{j}^{2}}$.
    \label{fig:example}}
\end{figure}

We then follow the method given in \cite{2014ApJ...792...54S} to
select epoch-pairs with significant ($\geqslant2\sigma$) variation in  space.
Here the variation is depicted as the distance
between two observations of an epoch-pair in $ {\rm log}F_{\rm X} - {\rm log}F_{\rm UV} $ space (see Fig. \ref{fig:example}).
The variability uncertainty $\sigma$ in $ {\rm log}F_{\rm X} - {\rm log}F_{\rm UV} $ space is computed as shown in Fig.~\ref{fig:example}. 

\begin{figure}
    \includegraphics[width=\columnwidth]{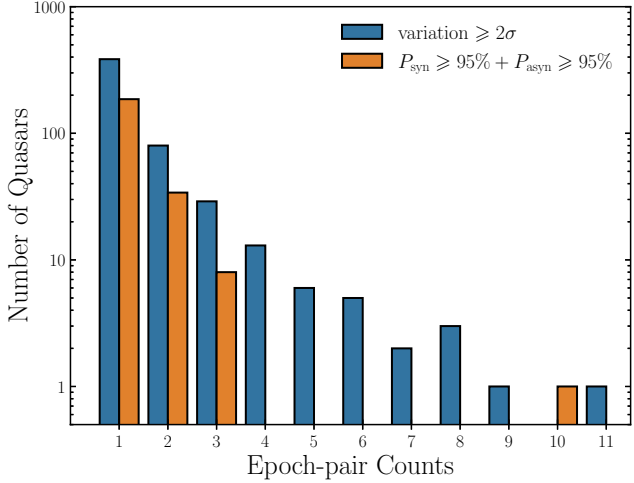}
    \caption{
        Histogram of the epoch-pair count for the 802 pairs of 525 quasars (with variation in $ {\rm log}F_{\rm X} - {\rm log}F_{\rm UV} $ space $\geqslant$ 2$\sigma$, blue histogram)
        and the $P_{\rm syn}$ or $P_{\rm asyn}$ $\geqslant 95\%$ epoch-pair sample (229 quasars, 288 pairs, orange histogram). 
        \label{fig:epoch_counts}}
\end{figure}

In this work
we only keep pairs with variation in $ {\rm log}F_{\rm X} - {\rm log}F_{\rm UV} $ space $\geqslant$ 2$\sigma$, and drop the rest for which the observed
variation could be dominated by photometric noise. 
We finally obtain a sample of 525 quasars with 802 epoch-pairs, 
with XMM exposure time of these epochs ranging from 0.017 to 1.65 days. Most (385) quasars have only one epoch-pair, and the largest number of epoch-pairs for a single quasar is 11 (PG 1114+445).  
In Fig.~\ref{fig:epoch_counts} we plot the number of quasars versus the corresponding number of epoch-pairs. 

\begin{figure}
    \includegraphics[width=\columnwidth]{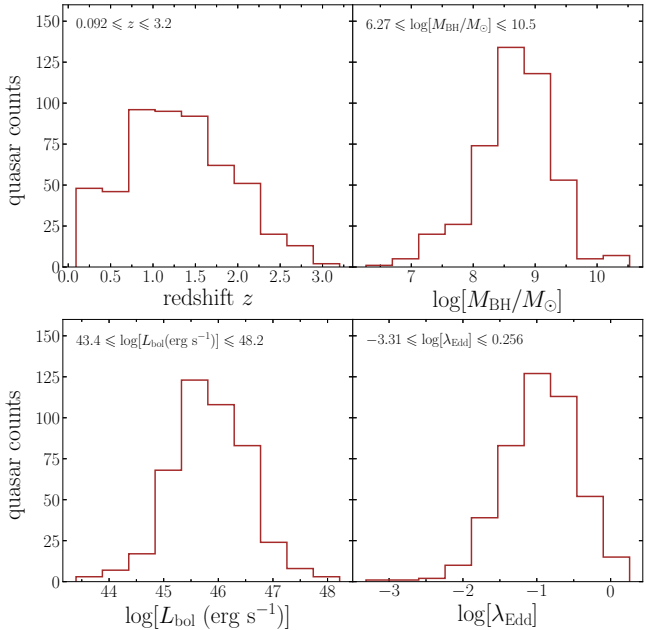}
    \caption{ Histograms of redshift, SMBH mass, bolometric luminosity, and Eddington ratio of our sample. 
        \label{fig:QSO_parameters}}
\end{figure}

We derive the time lag $\Delta t$ ($t_j$-$t_i$) for each epoch-pair, where $t_j$ or $t_i$ refers to the mid-point of the corresponding XMM exposure. The measurements of redshift, SMBH mass, bolometric luminosity and Eddington ratio for each quasar are taken from \cite{2020ApJS..249...17R}, and the distributions and ranges of these physical parameters of this sample are presented in Fig.~\ref{fig:QSO_parameters}. To explore later whether the relation between X-ray and UV variability depends on the relative X-ray brightness of quasars, 
we also calculate 
$\alpha_{\rm ox}$ which represents the ratio of the X-ray to UV luminosity for each quasar,
\begin{equation}
    \label{eqn:1}
    \alpha_{\rm ox} = \frac{{\rm log}(L_{\rm 2~keV}/L_{\rm 2500~\text{\AA}})}{{\rm log}(\nu_{\rm 2~keV}/\nu_{\rm 2500~\text{\AA}})}=0.3838\,{\rm log}(\frac{L_{\rm 2~keV}}{L_{\rm 2500~\text{\AA}}})
\end{equation}
where ${\rm log}(\nu)$ represents the logarithmic frequency at 2~keV or 2500~\text{\AA}. We calculate the 2~keV luminosity $L_{\rm 2~keV}$ 
from the observed 2 -- 4.5~keV band integral 
assuming a powerlaw spectrum with photon index $\Gamma = 1.7$ \citep{1997ApJ...477...93L}\footnote{Adopting $\Gamma$ = 1.9 would yield slightly larger $L_{\rm 2~keV}$ (by $\sim$ 0.1 dex on average), but would not change the major results of this work. }. 
The 2 -- 4.5~keV band was chosen in the favor of lesser absorption comparing to the lower energy band.
We adopt the method of \cite{2010A&A...519A..17V} to derive the rest-frame UV 2500~\AA~monochromatic luminosity $L_{\rm 2500~\text{\AA}}$.
If the spectral energy distribution (SED) obtained with the XMM OM filter set of UVW2 (1894~\AA), UVM2 (2205~\AA), UVW1 (2675~\AA),
U (3275~\AA), B (4050~\AA), and V (5235~\AA), covers the rest-frame 2500~\AA, 
we derive $L_{\rm 2500~\text{\AA}}$ through interpolating the two nearest (in wavelength) SED data points. If not, the extrapolation was
adopted by matching an average reference SED from \cite{2006ApJS..166..470R} to the nearest (in wavelength) SED data point. 
For an epoch-pair, we average the obtained $\alpha_{\rm ox}$ over two epochs for the following analyses.

\section{Results}
\label{sec:result}

\begin{figure}
    \includegraphics[width=\columnwidth]{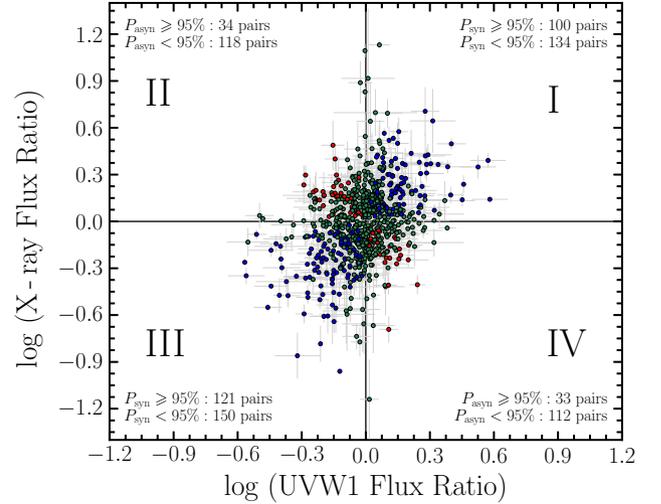}
    \caption{$\Delta{\rm log}F_{\rm X}$ versus $\Delta{\rm log}F_{\rm UV}$ for 802 epoch-pairs 
    with $\geqslant$ 2$\sigma$ variation in $ {\rm log}F_{\rm X} - {\rm log}F_{\rm UV} $ space (see Fig. \ref{fig:example}).
    Blue dots mark pairs with $P_{\rm syn}$ $\geqslant 95\%$ (in quadrant I and III), 
    i.e., pairs with clear synchronicity between X-ray and UV variability, while red points plot pairs with $P_{\rm asyn}$ $\geqslant 95\%$ (quadrant II and IV).
    ``Other pairs'' are plotted as green points. The numbers of pairs in each quadrant are also marked. 
    \label{fig:delta_flux}}
\end{figure}

We first explore whether the X-ray and UV variability of our quasars are coordinated, i.e., whether the brightening/dimming in X-ray and UV are synchronous. 
Note as for most of our quasars there is only one epoch-pair available, we are unable to study the coordination between X-ray and UV variability with cross-correlation function  \citep[e.g.,][]{1999PASP..111.1347W,2020MNRAS.495.1403X}. 
Instead, we calculate $\Delta{\rm log}F_{\rm UV}$ and $\Delta{\rm log}F_{\rm X}$ for each epoch-pair (the latter epoch - the earlier epoch) using the 0.5 -- 4.5 keV and UVW1 flux, and plot them on Fig. \ref{fig:delta_flux}. The data points fall in quadrant I or III in Fig. \ref{fig:delta_flux} implies the X-ray and UV variability are synchronous, while lying in quadrant II or IV signifies asynchronous variability (i.e., brightening in one band, but dimming in another one). 
Simply counting the number of epoch-pairs in each quadrant, we find 505 ($f_{\rm syn}$ = $63.0\pm1.7\%$\footnote{standard binomial error})
epoch-pairs with synchronous X-ray and UV variability, and 297 ($f_{\rm asyn}$ = $37.0\pm1.7\%$) pairs with asynchronous variability. Note that as there is no absolute ``rest'', we assume the intrinsic $\Delta{\rm log}F_{\rm UV}$ and $\Delta{\rm log}F_{\rm X}$ are always nonzero, and thus 
simply consider only two situations (synchronous or asynchronous variability), although for some epoch-pairs their variability in X-ray or UV could be insignificant compared with photometric noise. The effect of photometric noise on the observed $f_{\rm asyn}$ would be further discussed below and in \S\ref{sec:discussion}.

However, note that even if the X-ray and UV variations in all epoch-pairs are intrinsically well-coordinated (synchronous), the statistical errors in $\Delta{\rm log}F_{\rm UV}$ and $\Delta{\rm log}F_{\rm X}$ due to photometric noise could randomly place some epoch-pairs in quadrant II or IV.
For many epoch-pairs there are clear uncertainties in the synchronicity of their variability, 
particularly those lying close to the horizontal or vertical axes in Fig. \ref{fig:delta_flux} (i.e., with clear variation in one band but weak variation in the other). 
To quantify the confidence of variation synchronicity of an epoch-pair, we first calculate the signal to noise ratio of its variation in X-ray and UV band respectively, 
\begin{equation}
    {\rm S/N} = \left| {\rm log}F_{i}- {\rm log}F_{j}\right|/ \sqrt{\delta_{i}^{2} + \delta_{j}^{2}}
\end{equation}
where $F_{i,j}$ and $\delta_{i,j}$ denote the observed fluxes and uncertainties (in logarithm space)
of two epochs $t_{i}$ and $t_{j}$ ($t_{j} > t_{i}$)
respectively. We then compute the statistical probability that an epoch-pair has synchronous or asynchronous variability in X-ray and UV. For instance, for an epoch-pair brightening
in X-ray with S/N = 2, and dimming in UV with S/N = 1, 
its brightening probability in X-ray is $P_{\rm X, brighten} = 97.7\%$, and
dimming probability in UV is $P_{\rm UV, dim} = 84.1\%$ (derived from the cumulative distribution function of normal distribution). 
We then calculate the probability of asynchronous variability as
\begin{equation}
    \label{eqn:3}
    P_{\rm asyn} = 	P_{\rm X,brighten}P_{\rm UV,dim} + (1-P_{\rm X, brighten})(1-P_{\rm UV, dim})
\end{equation}
and probability of synchronous variability as $P_{\rm syn}$ = $1 - P_{\rm asyn}$.
For 288 epoch-pairs of 229 quasars, we find $P_{\rm syn}$ or $P_{\rm asyn}$ $\geqslant 95\%$, 
including 221 synchronous pairs (hereafter ``$P_{\rm syn}$ $\geqslant 95\%$ pairs''), 
and 67 asynchronous pairs (hereafter ``$P_{\rm asyn}$ $\geqslant 95\%$ pairs''). 
The remaining 514 pairs are named ``other pairs'' hereafter. 
If we focus on only these 288 pairs, the $f_{\rm asyn}$ is $23.3\pm2.5\%$ ($f_{\rm syn}$ = $76.7\pm2.5\%$).
Note this is a conservative lower limit to the asynchronous fraction, 
as the apparent $f_{\rm asyn}$ is obviously higher in ``other pairs'' (44.7\%).

\begin{figure}
    \includegraphics[width=\columnwidth]{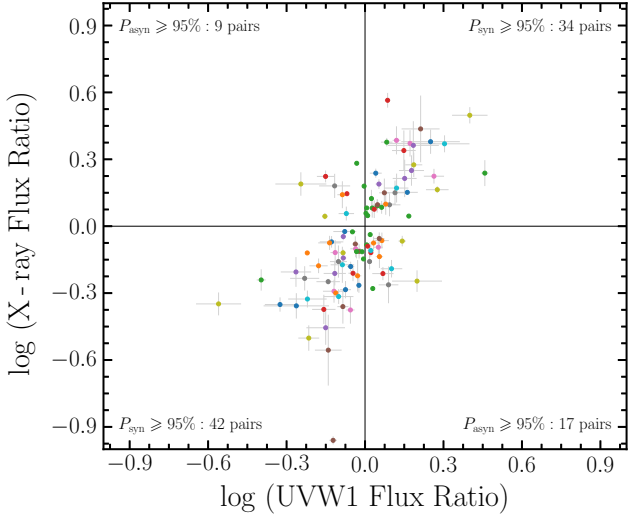}
    \caption{Same as Figure~\ref{fig:delta_flux} but for 43 quasars with multiple epoch-pairs with $P_{\rm syn}$ or $P_{\rm asyn}$ $\geqslant 95\%$. 
    Different quasars were marked with different 
    colors. 
    \label{fig:multiple pair}}
\end{figure}

There are 140 quasars (417 pairs) with more than one epoch-pairs, 76 quasars (267 pairs) consist of both synchronous and asynchronous variability, indicating that even in the same individual source, the X-ray and UV variability could switch from synchronous to asynchronous or vice versa. The asynchronous fraction of these 417 pairs is $38.1\%$, similar to 37.0\% derived from the full sample (802 pairs). 
If we only consider the pairs with $P_{\rm syn}\geqslant95\%$ or $P_{\rm asyn}\geqslant95\%$, there are 43 quasars consisting of more than one epoch-pairs. A total of 102 pairs with $P_{\rm syn} \geqslant95\%$ or $P_{\rm asyn} \geqslant 95\%$, with an $f_{\rm asyn}$ of 25.5\%, also similar to 23.3\% derived from all 288 pairs with $P_{\rm syn}$ $\geqslant$ 95\% or $P_{\rm asyn}$ $\geqslant$ 95\%. 
We plot these 102 pairs of the 43 quasars in Fig. \ref{fig:multiple pair}, 
which show distribution similar to those $P_{\rm syn}\geqslant95\%$ or $P_{\rm asyn}\geqslant95\%$ pairs in Fig. \ref{fig:delta_flux} 
(e.g., with similar correlation coefficient $\rho$ between the two quantities plotted, $\rho$ = 0.62 for Fig. \ref{fig:multiple pair}, and 0.57 for $P_{\rm syn} \geqslant95\%$ and $P_{\rm asyn} \geqslant95\%$ pairs in Fig. \ref{fig:delta_flux}).

Among the 43 quasars, 10 quasars have both $P_{\rm syn} \geqslant95\%$  and $P_{\rm asyn}$ $\geqslant$ 95\% pairs, exhibiting switches between synchronous and asynchronous variability in individual sources. 

\begin{figure*}
    \includegraphics[width=18cm,height=18cm]{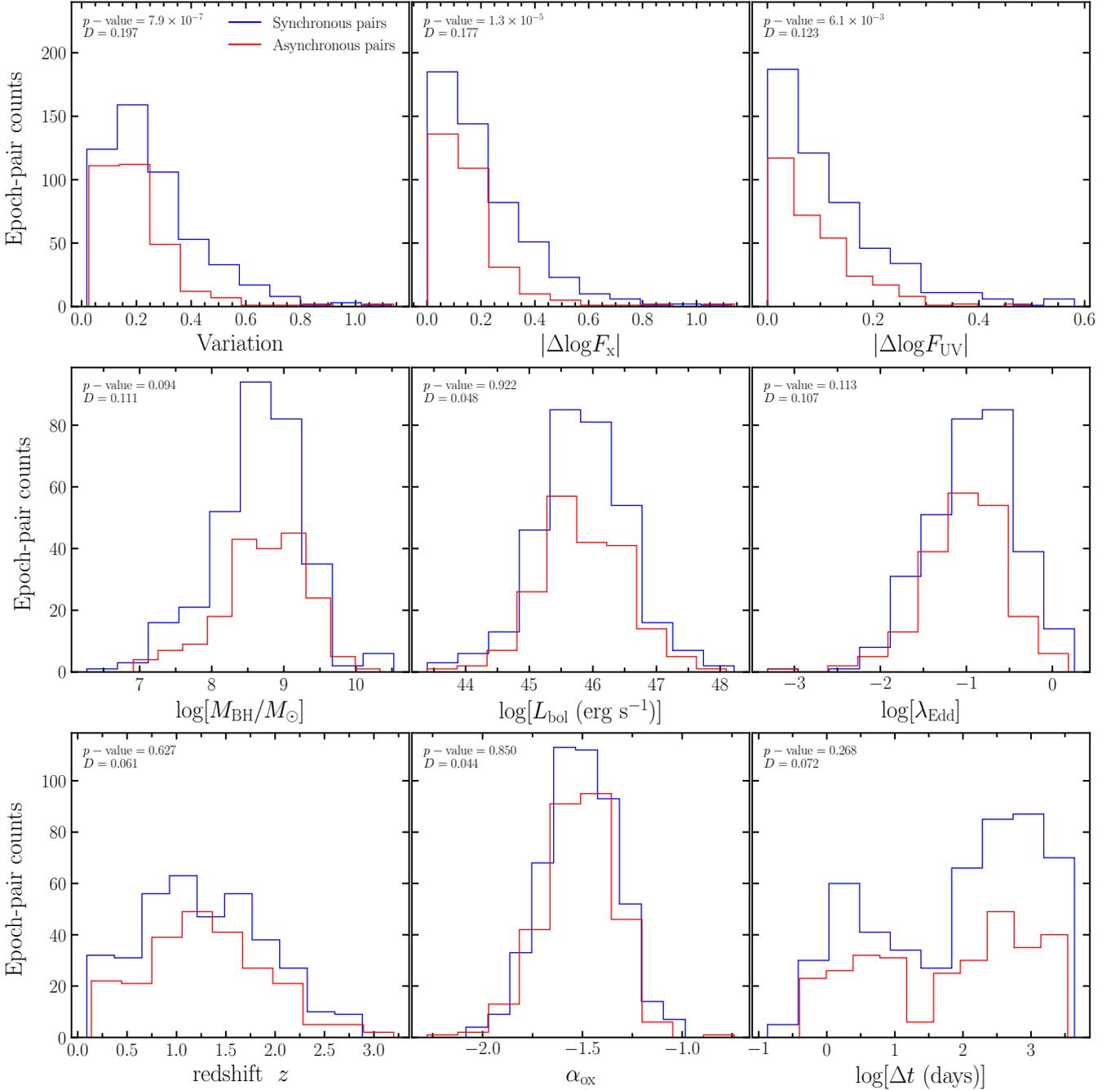}
    \caption{Histograms of observational and physical parameters of synchronous (quadrant I \& III in Fig. \ref{fig:delta_flux}) and asynchronous (quadrant II \& IV) pairs. The most upper-left panel plot the variation of an epoch-pair in $ {\rm log}F_{\rm X} - {\rm log}F_{\rm UV} $ space, i.e., $\sqrt{(\Delta {\rm log}F_{\rm X})^2+(\Delta {\rm log}F_{\rm UV})^2}$. In each histogram, the K-S test results between the two subsamples are labeled. For parameters such as redshift $z$, $M_{\rm BH}$, $L_{\rm bol}$ and $\lambda_{\rm Edd}$, a single quasar may contribute multiple epoch-pairs to a subsample. In that case, the quasar is counted once in the corresponding histogram.} \label{fig:properties_correlation}
\end{figure*}

We then explore whether the synchronicity of variability relies on observational or physical parameters of quasars, including variability amplitude, redshift, SMBH mass, bolometric luminosity, Eddington ratio, $\alpha_{\rm ox}$, and rest-frame time lag (see Fig. \ref{fig:properties_correlation}). We find that synchronous pairs have significantly larger variability amplitude than asynchronous pairs, indicating stronger variability is more likely synchronous between X-ray and UV.
Other than that, we find no statistical difference between the parameter distributions of synchronous and asynchronous pairs.

\begin{figure}
	\includegraphics[width=\columnwidth]{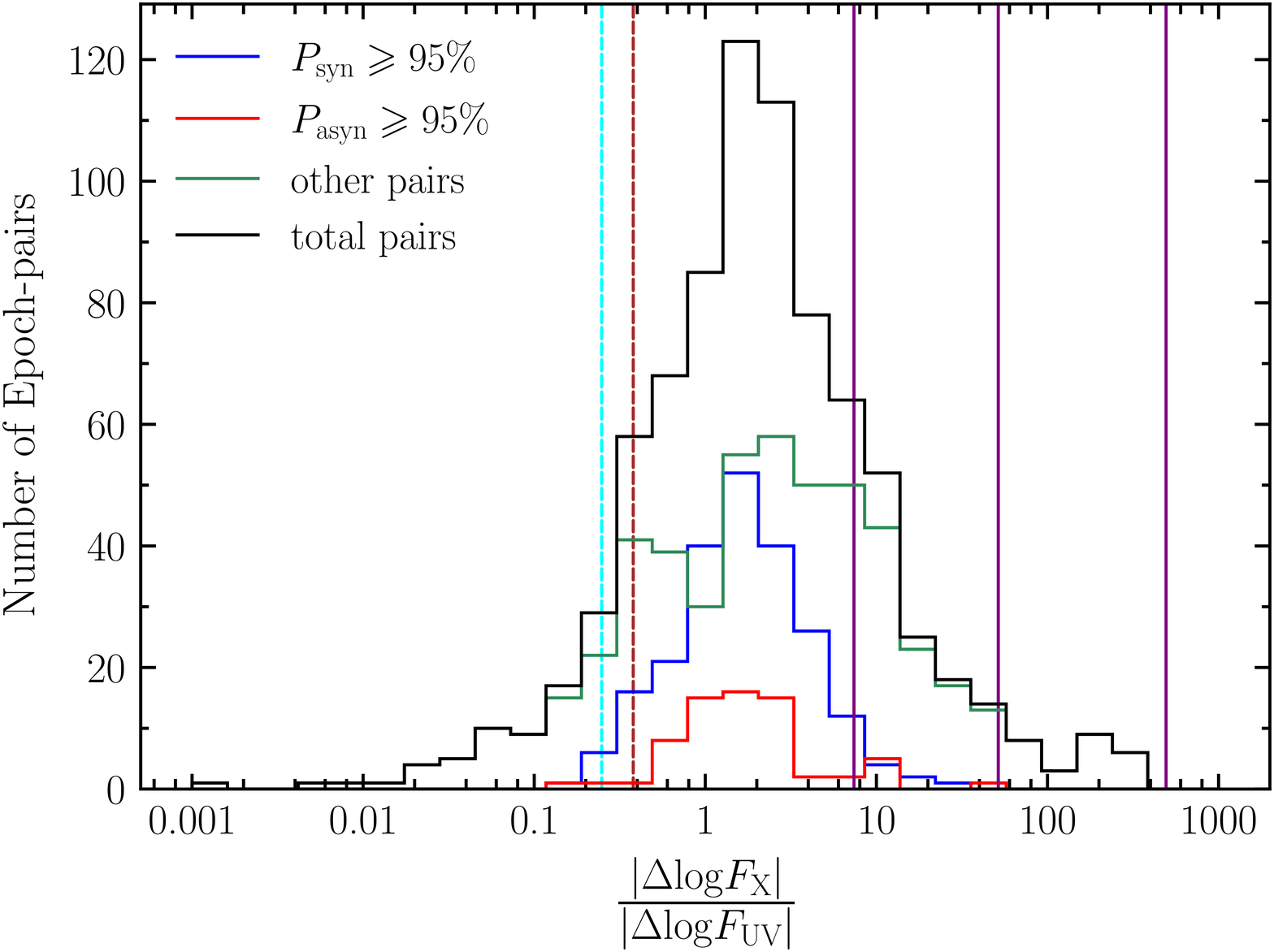}
    \caption{Distribution of the ratio of X-ray variability amplitude to UV variability amplitude.
        The blue and red histograms plot the epoch-pairs with $P_{\rm syn} \geqslant 95\%$ and $P_{\rm asyn} \geqslant 95\%$, 
        while the green histogram shows the epoch-pairs below the confidence of $95\%$.
        Two vertical dashed lines on the left mark the expected ratios due to changes of accretion rate, 
        derived based on the the type-1 AGN sample (cyan) and type-2 sample (brown) of  \protect\cite{2012MNRAS.425..623L}. 
        Three purple lines on the right represent the simulated X-ray to UV variability amplitude ratios 
        assuming different values of $L_{\rm X}/L_{\rm disk}$ = 1, 0.1, 0.01 
        (from left to right) in the reprocessing scenario (see text for details).}
        \label{fig:amplitude_ratio}
\end{figure}

To compare the X-ray and UV variability amplitudes ($|\Delta{\rm log}F_{\rm X}|$ and $|\Delta{\rm log}F_{\rm UV}|$) derived for each pair, we plot in Fig. \ref{fig:amplitude_ratio} the histogram of $|\Delta{\rm log}F_{\rm X}|$/$|\Delta{\rm log}F_{\rm UV}|$. 
We find 565 ($70.4\pm1.6\%$) of all 802 pairs have larger X-ray variability amplitude than in UV. 
The number is 212 ($73.6\pm2.6\%$) if considering only pairs with $P_{\rm syn}$ $\geqslant$ 95\% ($P_{\rm asyn}$ $\geqslant$ 95\%).
Meanwhile, $|\Delta{\rm log}F_{\rm X}|$/$|\Delta{\rm log}F_{\rm UV}|$ spans a very broad range (over 5 dex), which may be partly attributed to photometric noise. 

\begin{figure*}
    \includegraphics[width=18cm, height=18cm]{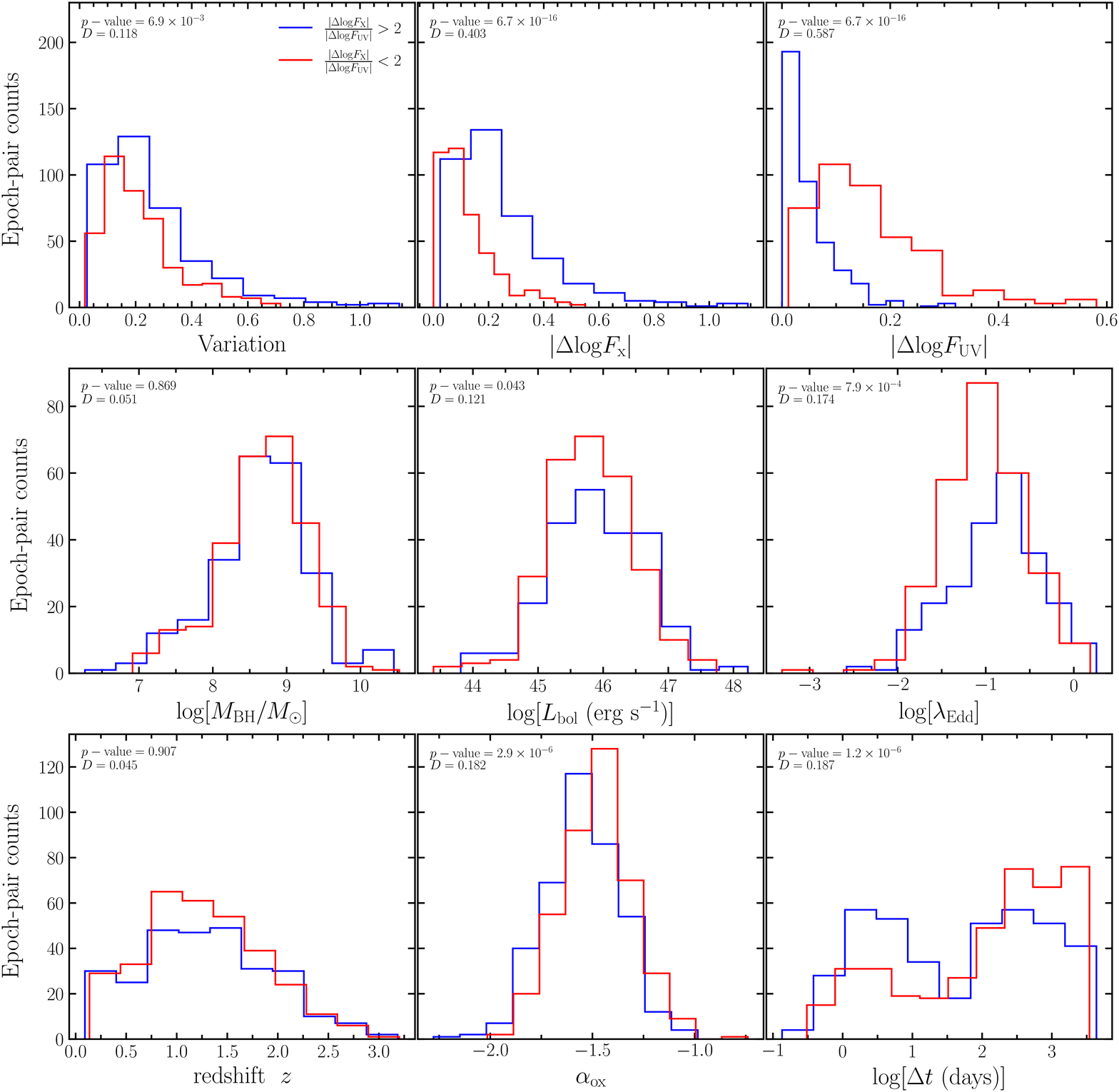}
    \caption{Similar to Fig. \ref{fig:properties_correlation}, 
    but here we present the histograms for two subsamples split according 
    to $|\Delta{\rm log}F_{\rm X}|$/$|\Delta{\rm log}F_{\rm UV}|$.
    Note in the upper middle and right panel, 
    the comparison has no physical meaning, 
    as pairs with larger $|\Delta{\rm log}F_{\rm X}|$ ($|\Delta{\rm log}F_{\rm UV}|$) 
    automatically have larger (smaller) $|\Delta{\rm log}F_{\rm X}|$/$|\Delta{\rm log}F_{\rm UV}|$.}
    \label{fig: properties_amplitude}
\end{figure*}

We finally explore the dependency of $|\Delta{\rm log}F_{\rm X}|$/$|\Delta{\rm log}F_{\rm UV}|$ on observational and physical parameters of quasars (see Fig. \ref{fig: properties_amplitude}). We split the 802 pairs into two subsamples according to $|\Delta{\rm log}F_{\rm X}|$/$|\Delta{\rm log}F_{\rm UV}|$ ($>2$ vs $<2$), and find pairs with larger $|\Delta{\rm log}F_{\rm X}|$/$|\Delta{\rm log}F_{\rm UV}|$ tend to have stronger variation (in $ {\rm log}F_{\rm X} - {\rm log}F_{\rm UV} $ space), higher Eddington ratio, relatively weaker X-ray emission (smaller $\alpha_{\rm ox}$), and shorter time lag.

\section{discussion} \label{sec:discussion}

\begin{figure}
    \includegraphics[width=\columnwidth]{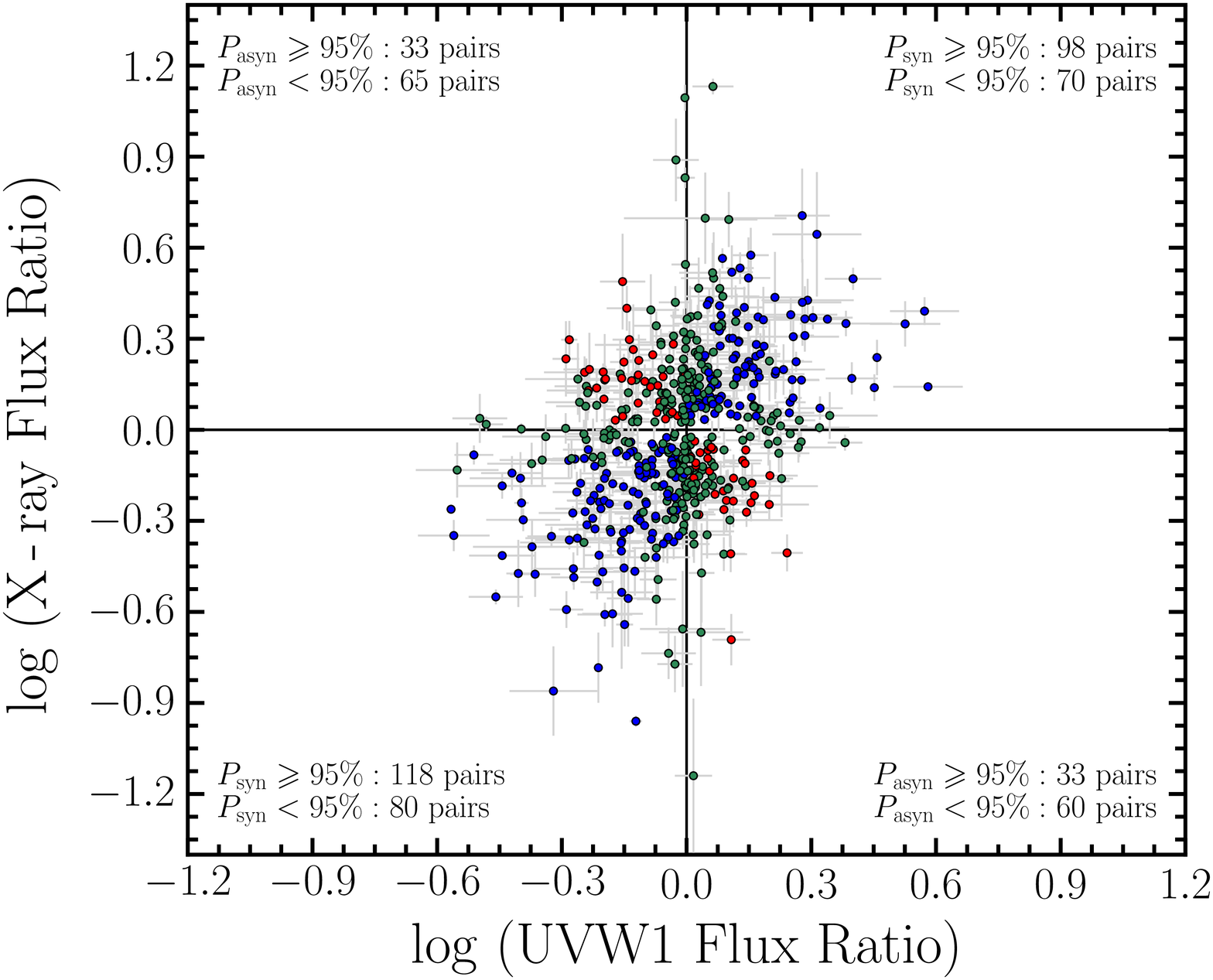}
    \caption{Same as in Fig. \ref{fig:delta_flux} but only for the epoch-pairs with $\geqslant$ 3$\sigma$ variation in $ {\rm log}F_{\rm X} - {\rm log}F_{\rm UV} $ space.
        \label{fig:3_sigma}}
\end{figure}

We have shown that the X-ray/UV variability of quasars is clearly asynchronous in a remarkable fraction of epoch-pairs observed with XMM-Newton. 
Utilizing epoch-pairs with $\geqslant$ 2$\sigma$ variation in $ {\rm log}F_{\rm X} - {\rm log}F_{\rm UV} $ space, 
we find an asynchronous fraction $f_{\rm asyn}$ of $37.0\pm1.7\%$, derived directly from counting the fraction of pairs in quadrant II and IV.
If only counting pairs with $P_{\rm syn}$ or $P_{\rm asyn}$ $\geqslant 95\%$, we obtain a conservative lower limit to $f_{\rm asyn}$ of $23.3\pm2.5\%$.
We further confirm the results by using only epoch-pairs with $\geqslant 3\sigma$ variation in $ {\rm log}F_{\rm X} - {\rm log}F_{\rm UV} $ space (see Fig. 
\ref{fig:3_sigma}), as the effect of noise may still be non-negligible for variations with significance between 2-3$\sigma$. 
Repeating our analyses, and from Fig. \ref{fig:3_sigma} 
we obtain an asynchronous fraction of $34.3\pm2.0\%$ (and $23.4\pm2.5\%$ if only counting pairs with $P_{\rm syn}$ or $P_{\rm asyn}$ $\geqslant 95\%$). 

We further perform Monte-Carlo simulations to quantify the effect of photometric noise on the directly derived $f_{\rm asyn}$. 
We add random Gaussian photometric errors to data points in Fig. \ref{fig:delta_flux} and repeating 10,000 times to obtain 10,000 groups of 802 pairs. For each group we re-calculate its $f_{\rm asyn}$. 
The mean value of the 10,000 simulated $f_{\rm asyn}$ is 38.4\%, with a standard deviation of $\pm1.2\%$. Note the scatter to simulated $f_{\rm asyn}$ (1.2\%) caused by photometric noise is even smaller than the standard binomial error of the observed $f_{\rm asyn}$ (1.7\%). This indicates photometric errors have only minor effect on the observed $f_{\rm asyn}$, by slightly increasing the input $f_{\rm asyn}$ by 1.4\%.
Therefore, in reverse we expect the intrinsic $f_{\rm asyn}$ of the observed 802 pairs to be $37.0\% -1.4\%$ = $35.6\pm2.1\%$.

The fraction of asynchronous variability is remarkably high. 
Assuming X-ray and UV variations in quasars are mutually independent, 
we would expect the same number of epoch-pairs in quadrant I and III, 
as those in quadrant II and IV. Thus an asynchronous fraction of A\% indicates in 2$\times$A\% of the epoch-pairs the X-ray and UV variations are completely independent, and only $1-2\times$A\% of epoch-pairs are intrinsically coordinated. Therefore the asynchronous fraction of $35.6\pm2.1\%$ indeed indicates in only $28.8\pm4.2\%$ of the epoch-pairs the X-ray and UV variability are intrinsically coordinated. 

We note that variable X-ray absorbers (e.g. from disc winds,  warm absorbers, etc) along the line of sight may be responsible for X-ray flux variation in some AGNs \citep[e.g.,][]{2009A&ARv..17...47T, 2014MNRAS.439.1403M}. Such absorbers, if dust-free and/or with too small scale to block the UV emitting region, could attenuate only (mainly soft) X-ray fluxes but not UV, and thus could yield X-ray variability uncoordinated with UV variation. Note that the median redshift of our sample is 1.25, thus the observed 0.5 -- 4.5 keV corresponds to the 1.12 -- 10.1 keV in the quasar rest-frame, indeed less insensitive to X-ray absorption compared with $<$ 1 keV band in rest frame. 
Furthermore, our sample only includes type-1 quasars for which the X-ray absorption, if exists, is expected to be weak for most sources.

We also repeat our analyses using the observed 2 -- 4.5 keV (instead of 0.5 -- 4.5 keV) fluxes, and derive a rather similar $f_{\rm asyn}$ of $38.3\pm1.9\%$ (compared with $37.0\pm1.7\%$ using 0.5 -- 4.5 keV fluxes), and $f_{\rm asyn}$ = $27.2\pm3.5\%$ if considering only $P_{\rm syn}\geqslant 95\%$ or $P_{\rm asyn}\geqslant95\%$ pairs (also very close to $23.3\pm2.5\%$ using 0.5 -- 4.5 keV fluxes). Therefore, variable X-ray absorption could unlikely play a significant role in biasing the results of this work. 

\subsection{Challenging the X-ray reprocessing model and accretion rate variation model}

Our discovery that in a dominant fraction of the epoch-pairs quasars show intrinsically uncoordinated X-ray and UV variability clearly challenge the X-ray reprocessing model and the scenario of accretion rate change for variation in quasars, as both models predict well-coordinated variations between X-ray and UV.  

\begin{figure}
    \includegraphics[width=\columnwidth]{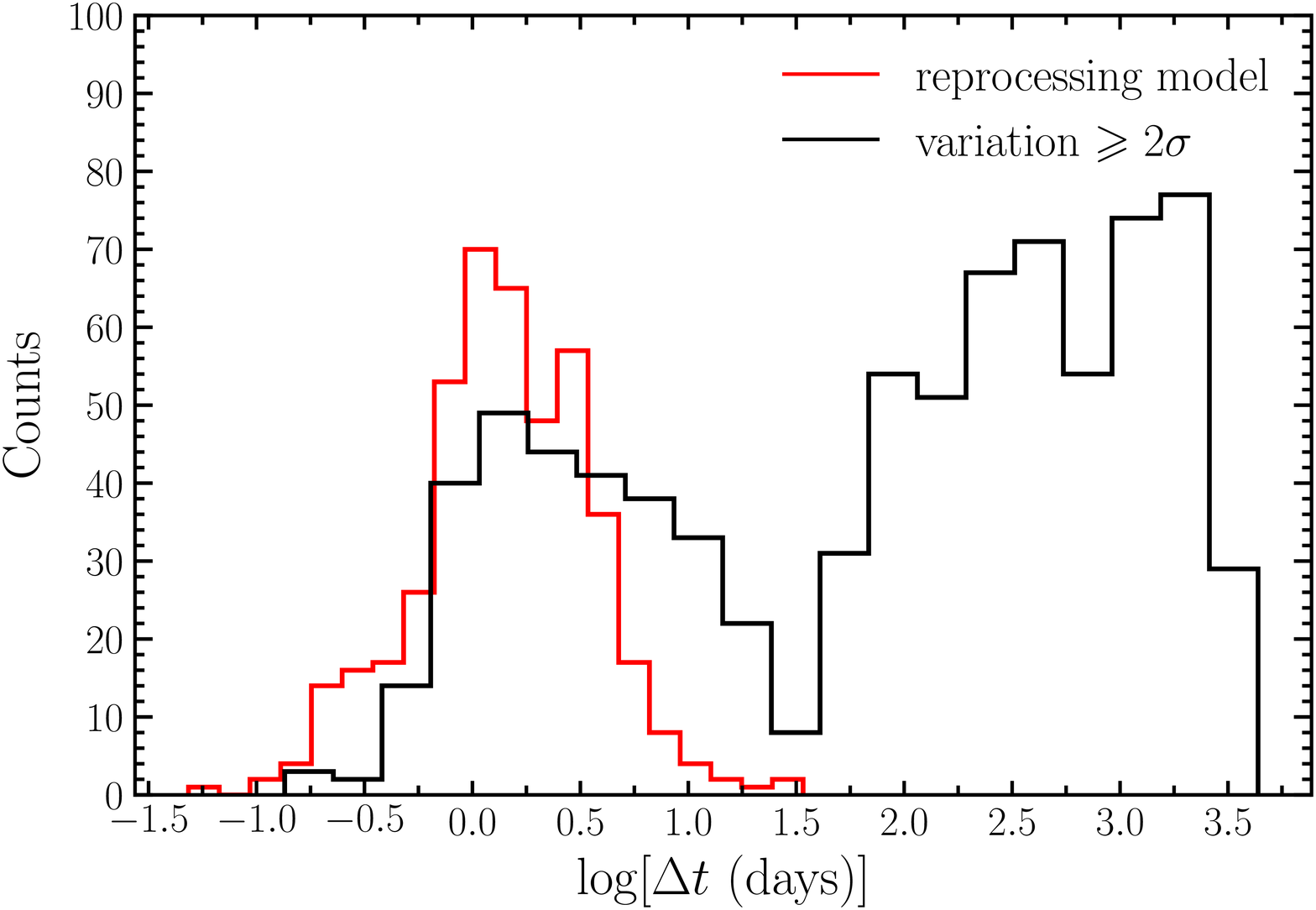}
    \caption{The histogram of rest-frame time lags of 802 epoch-pairs analysed in this work (black), compared with the expected X-ray/UV lag in the reprocessing scheme (red), computed following Equation (12) of \citep{2016ApJ...821...56F}, i.e., 
    the light travel time from the central variable source to the reprocessed UV region ($2500~\text{\AA}$).
    \label{fig:timescale}}
\end{figure}

In the reprocessing scheme a time delay of $\sim$ days for quasars is expected between X-ray and UV/optical variability. Observations also detected time delays between the X-ray and UV/optical band for several quasars, ranging from days to weeks  \citep{2008MNRAS.389.1479A, 2009MNRAS.399..750B, 2009ApJ...696..601M}. 
We compute this expected reprocessing time lag (between X-ray and 2500~\AA) for our quasars, using the measured SMBH mass and Eddington ratio, and 
adopting the Equation (12) of \cite{2016ApJ...821...56F}.
Fig. \ref{fig:timescale} shows that the expected reprocessing time lags are generally much smaller than the observed lags of the epoch-pairs we utilized, thus the lag between X-ray and UV variability is unlikely to affect the coordination analyses in this work. 
Meanwhile, the synchronicity of X-ray/UV variability shows no dependence on observed lag (see Fig. \ref{fig:properties_correlation}), also suggesting our analyses is not affected by the potential lag between UV and X-ray variability in our quasars. 

Comparing the variation amplitude in X-ray and UV could also provide essential constraints to physical models. 
For comparison, we estimate the ratio of X-ray/UV variability amplitude in the framework of reprocessing and accretion rate variation respectively. 
For the reprocessing model, we apply the method in \cite{2018ApJ...860...29Z} to generate the X-ray/UV light curves observed by the UVW1 filter of {\em XMM-Newton}. 
We adopt the necessary parameters in Table 1 of \cite{2018ApJ...860...29Z},
a SMBH mass of ${\rm log}M_{\rm BH}\sim8.32$ and Eddington ratio of ${\rm log}\lambda_{\rm Edd}\sim -0.913$, typical values for our quasar sample. 
We conduct three simulations assuming $L_{\rm X}/L_{\rm disk}$ = 0.01, 0.1 and 1, respectively\footnote{Here $L_{\rm X}$ is simply considered as input energy, in addition to the energy released in the standard disk via accretion.}. As a result, 
even in the most extreme case ($L_{\rm X}/L_{\rm disk}$ = 1), 
the observed X-ray/UV variability amplitude ratios in most pairs appear too small for the reprocessing model. 
Note in quasars the observed X-ray (2 -- 10~keV) to bolometric luminosity correction factor is generally 10-100 \citep[e.g.,][]{2012MNRAS.425..623L}, implying 
$L_{\rm 2-10~keV}/L_{\rm disk}$ is only $\sim$ 0.01-0.1.

For the accretion rate variation model, the $L_{\rm X}/L_{\rm bol}$ in samples of AGNs is known to be Eddington ratio dependent \citep[e.g.,][]{2012MNRAS.425..623L}.
Such dependence could be converted to $|\Delta{\rm log}F_{\rm X}|$/$|\Delta{\rm log}F_{\rm UV}|$, assuming variability in individual sources is simply due to changes of accretion rate. 
\cite{2012MNRAS.425..623L} reported $L_{\rm bol}$/$L_{\rm 2-10~keV}$ $\propto$ $\lambda_{\rm Edd}^{0.752}$ and $\lambda_{\rm Edd}^{0.621}$ (for type 1 and type 2 quasars, respectively, see Table 3 and Table 4 of \citealt{2012MNRAS.425..623L}), i.e., 
$L_{\rm X} \propto L_{\rm bol}^{0.248}$ and $L_{\rm X} \propto L_{\rm bol}^{0.379}$.
This indicates when Eddington ratio increases by one order of magnitude, the X-ray luminosity only increases by 0.248 or 0.379 dex.  
Since in quasars the bolometric luminosity is dominated by UV emission,
we would expect $|\Delta{\rm log}F_{\rm X}|$/$|\Delta{\rm log}F_{\rm UV}|$ of 0.248 and 0.379 (see Fig. \ref{fig:amplitude_ratio}) if the variability is simply due to changes of accretion rate.
The observed $|\Delta{\rm log}F_{\rm X}|$/$|\Delta{\rm log}F_{\rm UV}|$ of the dominant fraction of epoch-pairs are considerably larger, thus also contradicting the accretion rate variation model.

Clearly, the results presented in this work significantly challenge both the X-ray reprocessing mode 
and accretion rate variation model from two aspects: 
1) the coordination between X-ray and UV variability in our quasar sample is poor; 
2) the observed amplitude ratios of X-ray and UV variability appear too small for reprocessing model, 
while too large for accretion rate variation model.
Note in literature there already exist other challenges to these two models. 
A (likely incomplete) list includes, 1) poor correlation between X-ray and UV/optical reported in individual AGNs (see \S\ref{sec:intro}); 
2) the UV to X-ray time lags detected in several Seyfert galaxies were larger 
than the prediction of the reprocessing scheme 
\citep[e.g.,][]{2015ApJ...806..129E, 2016AN....337..500M, 2017ApJ...840...41E, 2018MNRAS.480.2881M}; 3) the UV/optical variability in NGC 5548 appears too smooth to be reprocessed by X-ray \citep{Gardner2017}; 4) the amplitude of UV/optical color variability in quasars is inconsistent with accretion rate changes \citep{Schmidt2012}; 5) the timescale dependence of the color variability of quasars disfavor both accretion rate changes \citep{2014ApJ...792...54S,ZhuFF2016,Cai2019} and reprocessing \citep{2018ApJ...860...29Z}; 6) the X-ray spectral variability in individual AGNs is quantitatively inconsistent with accretion rate change \citep{Sobolewska2009} and timescale dependent \citep{Wu2020}; and etc. 

\subsection{The nature of the relation between X-ray and UV variability}

The partial coordination between UV and X-ray variability in quasars we find in this work
implies X-ray and UV variability in quasars are not completely independent of each other. They are somehow related, but not as strong as the simple X-ray reprocessing model or accretion rate variation model predicts. 
The observed coordination (synchronicity) exhibits no dependence on physical parameters including time lag, redshift, SMBH mass, bolometric luminosity, Eddington ratio, and $\alpha_{\rm ox}$, except for that stronger variations tend to be better correlated between X-ray and UV. 
Meanwhile, both correlated and uncorrelated variations are similarly seen in individual sources. 
These facts require underlying mechanism(s) which could produce both correlated and uncorrelated variability between X-ray and UV. 

Such observational results clearly favor the inhomogeneous disk model for AGN variations. In the original inhomogeneous disk model \citep{2011ApJ...727L..24D, 2016ApJ...826....7C}, the quasar variability in UV/optical is considered to originate from temperature fluctuations in individual zones of the accretion disk,
though expected to be poorly correlated between different wavelengths \citep[e.g.,][]{10.1093/mnras/stv241}. 
In a revised inhomogeneous disk model proposed by \cite{2018ApJ...855..117C}, a common large scale fluctuation (to mimic the effect of turbulence propagation along all directions in the disk) is superposed on the independent local fluctuation. The revised model could reproduce tight inter-band correlations between UV/optical bands and simultaneously reproduce continuum lags between various bands, both of which well match the observations of NGC 5548 \citep{2018ApJ...855..117C}. Note in \cite{2018ApJ...855..117C}, the inter-band continuum lags are caused by different regression capability
of different disk regions when responding to the common large scale fluctuation, i.e., the inner hotter disk region could regress faster than outer cooler region, thus short-wavelength variability appears leading that of long-wavelength. 
\cite{2020ApJ...892...63C} further extended the model to incorporate X-ray variability, and found the model could explain the puzzling large UV to X-ray lags detected in several Seyfert galaxies.

The inhomogeneous disk model could naturally produce correlated and 
uncorrelated variability between X-ray and UV. 
In case of stronger variation we would expect stronger large scale fluctuation in the disk, 
thus stronger correlation between X-ray and UV variability is also expected. 
The stochastic nature of variability produced by the 
inhomogeneous disk model may also explain the broad range of the observed $|\Delta{\rm log}F_{\rm X}|$/$|\Delta{\rm log}F_{\rm UV}|$. 
Furthermore, the observed dependence of $|\Delta{\rm log}F_{\rm X}|$/$|\Delta{\rm log}F_{\rm UV}|$ on physical parameters reported in \S\ref{sec:result} 
may also be used to constrain the parameters of the inhomogeneous disk model.
Quantitative comparison between the model and our results however is beyond the scope of this work, and is postponed to a later study. 

The stochastic nature of the relation between variations in different bands could have important implications. For instance, it could affect the broad line reverberation mapping studies, as the ionizing continuum may be poorly correlated with the observed UV/optical continuum \citep{Goad2016,Gaskell2021}. 
We finally stress that in relevant studies, because of the randomness, observed results could often differ or even contradict from source to source, or even from observation to observation (or from epoch to epoch) of the same source. 
Longer monitoring campaigns and/or larger sample statistical studies are essential to tackle the stochastic nature of the relation between variations in different bands. 

\section*{Acknowledgements}
We thank the anonymous reviewers for their constructive comments, which helped us to improve the manuscript. 
This work was supported by the National Science Foundation of China (No. 1890693, 12033006 $\&$ 12192221). The authors gratefully acknowledge the support of Cyrus Chun Ying Tang Foundations.

 \section*{Data Availability}
The data underlying this article are available in public archives. The data reduction made use of the {\em XMM-Newton} Science Archive (XSA),
Astropy \citep{2013A&A...558A..33A}, Matplotlib \citep{2007CSE.....9...90H} 
and Topcat \citep{2005ASPC..347...29T}.
 



\bibliographystyle{mnras}
\bibliography{example} 








\bsp	
\label{lastpage}
\end{document}